\documentclass[conference]{IEEEtran}
\IEEEoverridecommandlockouts
\usepackage{cite}
\usepackage{amsmath,amssymb,amsfonts}
\usepackage{algorithmic}
\usepackage{graphicx}
\usepackage{textcomp}
\usepackage{xcolor}
\def\BibTeX{{\rm B\kern-.05em{\sc i\kern-.025em b}\kern-.08em
    T\kern-.1667em\lower.7ex\hbox{E}\kern-.125emX}}

\usepackage{changes}
\definechangesauthor[name="Ilya Safro", color=red]{is}
\definechangesauthor[name="Yuri Alexeev", color=blue]{ya}
\definechangesauthor[name="Cameron Ibrahim", color=purple]{ci}

\DeclareMathOperator{\incident}{inc}
\DeclareMathOperator{\degree}{deg}

\DeclareMathOperator{\shared}{share}

\DeclareMathOperator{\ops}{ops}
\DeclareMathOperator{\tsize}{size}
\DeclareMathOperator{\cost}{cost}

\newcommand{\Real}[0]{\mathbb{R}}

\usepackage{algorithm}
\usepackage{subcaption}

\usepackage{hyperref}

\begin{document}

\title{Constructing Optimal Contraction Trees for Tensor Network Quantum Circuit Simulation
}

\author{\IEEEauthorblockN{Cameron Ibrahim\IEEEauthorrefmark{2}, Danylo Lykov \IEEEauthorrefmark{1}, Zichang He \IEEEauthorrefmark{3}, Yuri Alexeev \IEEEauthorrefmark{1}, Ilya Safro\IEEEauthorrefmark{2}}
\IEEEauthorblockA{\IEEEauthorrefmark{1}Computational Science Division, Argonne National Laboratory, Argonne IL, USA}
\IEEEauthorblockA{\IEEEauthorrefmark{2}Department of Computer and Information Sciences, University of Delaware, Newark DE, USA}
\IEEEauthorblockA{\IEEEauthorrefmark{3}University of California Santa Barbara, Santa Barbara, California, USA}
}

\maketitle









 



\begin{abstract}
One of the key problems in tensor network based quantum circuit simulation is the construction of a contraction tree which minimizes the cost of the simulation, where the cost can be expressed in the number of operations as a proxy for the simulation running time. This same problem arises in a variety of application areas, such as combinatorial scientific computing,  marginalization in probabilistic graphical models, and solving constraint satisfaction problems. In this paper, we reduce the computationally hard portion of this problem to one of graph linear ordering, and demonstrate how existing approaches in this area can be utilized to achieve results up to several orders of magnitude better than existing state of the art methods for the same running time. To do so, we introduce a novel polynomial time algorithm for constructing an optimal contraction tree from a given order. Furthermore, we introduce a fast and high quality linear ordering solver, and demonstrate its applicability as a heuristic for providing orderings for contraction trees. Finally, we compare our solver with competing methods for constructing contraction trees in quantum circuit simulation on a collection of randomly generated Quantum Approximate Optimization Algorithm Max Cut circuits and show that our method achieves superior results on a majority of tested quantum circuits.\\
\noindent Reproducibility: Our source code and data are available at \url{https://github.com/cameton/HPEC2022_ContractionTrees}.
\end{abstract}



\section{Introduction}

Tensor networks have become a ubiquitous tool for describing high dimensional, data-sparse tensors in a space efficient manner\cite{Cichocki_2016, gray_hyperoptimized, robeva_graphical}. In quantum circuit simulation, tensor networks can be used to model circuits with far more qubits than a more direct approach such as state vector simulation, so long as the circuit is sufficiently simple \cite{lykov2021performance}. As such, tensor network based simulation methods will remain vital for the study of quantum algorithms that require a large number of qubits until sufficient advances are made with quantum hardware \cite{lykov2021performance}. Moreover, even with the advanced hardware, such simulation methods will be important for such tasks as finding suitable parameters in variational quantum algorithms that utilize both quantum and classical machines \cite{liu2022layer}. These networks arise naturally in a variety of disciplines such as Combinatorial Scientific Computing \cite{hendrickson2006combinatorial}, Probabilistic Modelling \cite{robeva_graphical}, Constraint Satisfaction \cite{Kourtis_2019}, Machine Learning and Data Mining \cite{efthymiou2019tensornetwork}, Physics Modelling \cite{montangero2018introduction}, and Quantum \& Classical Circuit Simulation\cite{gray_hyperoptimized, markov2008simulating}. 

A tensor network is most easily conceptualized as a collection of high-order matrices called tensors which are connected by edges. These networks are often used to represent tensors which are impractical to store explicitly in memory\cite{Cichocki_2016}. The fundamental operation performed on these networks is called a tensor contraction, where two tensors are combined to create a third tensor. 
 A series of tensor contractions performed on the network is called a tensor network contraction, which is used to reduce the overall size of the network in order to evaluate entries of the underlying tensor that the network represents.

It is useful to represent a tensor network contraction as a contraction tree, a binary tree whose leaves are tensors in the network, where each internal vertex represents the tensor contraction of its two children.
In the general case, the cost of any single tensor contraction is exponential in the number of dimensions of its inputs, as is the size of its output\cite{ogorman_parameterized}. 
As such, a vital part of performing a tensor network contraction is choosing a contraction tree which minimizes some cost, such as minimizing the size of any intermediary tensor or the total number of operations performed during your tensor network contraction\cite{gray_hyperoptimized}.
However, it is NP-hard to construct an optimal contraction tree in the general case for these objectives \cite{ogorman_parameterized}, meaning no optimal solution can been found deterministically in polynomial time. \\


\noindent {\bf Our contribution} 
In this paper, we will show how the computationally hard portion of constructing a contraction tree can be reduced to a problem of linear ordering, and give algorithms for constructing an optimal contraction tree for a given ordering. Furthermore, we will give a useful heuristic for choosing orderings which will result in high quality contraction trees. This algorithm is being developed with integration with the QTensor quantum circuit simulation library in mind \cite{qtensor,lykov2020tensor,lykov2021performance,lykov2021importance}. 
In general, in order to run simulations of quantum circuits with large numbers of qubits, we must be able to find a good quality contraction order. As such, finding optimal or near optimal contraction orders is critical for the development and testing of new quantum algorithms, benchmarking and profiling upcoming quantum devices, and verification of quantum advantage and supremacy claims to name a few applications \cite{alexeev2021quantum}
\emph{Even small improvements to the quality of the contraction tree will lead to a significant acceleration of quantum simulation. We demonstrate an improvement over competitive solvers by orders of magnitude on some instances for a comparable running time.}

The paper is structured the following way. Section \ref{sec:bn} provides an introduction to tensor networks and linear orderings. Section \ref{sec:related} introduces existing solvers used for constructing contraction trees for quantum circuit simulation, which we use for benchmarking the proposed algorithm. In Section \ref{sec:method}, we present a novel polynomial time algorithm for constructing an optimal contraction tree from a given order and explain the ordering heuristic that was used. In Section \ref{sec:results}, we evaluate the developed tree structure optimization algorithm against Tamaki-2017, FlowCutter, and Cotengra on a collection of QAOA circuits on different comparison metrics. Finally, Section \ref{sec:future} discusses possible areas for future work, while Section \ref{sec:conclusion} offers a conclusion.

\section{Background and notation}\label{sec:bn}

\subsection{Matrix Chains and Tensor Networks}

An \(m\) by \(n\) matrix \(M\) is a 2 dimensional array of scalar numbers from some field \(\mathbb{F}\) (e.g. \(\mathbb{R}\), \(\mathbb{C}\)). We say this matrix has two indices, \(i\) and \(j\), where \(\tsize(i) = m\) and \(\tsize(j) = n\). The size of \(M\) is \(\tsize(M) = \tsize(i)\tsize(j)\).

A matrix \(M_1\) with indices \(\{i,j\}\) and a matrix \(M_2\) with indices \(\{j,k\}\) can be combined to form a matrix \(M_1 \ast M_2\) with indices \(\{i,k\}\) using an operation known as matrix multiplication. Note that the output has the indices of both its inputs with the shared index removed. The number of operations required to perform a matrix multiplication is \(\tsize(i)\tsize(j)\tsize(k)\).

\begin{figure}[ht]
    \centering
    \includegraphics[width=0.30\textwidth]{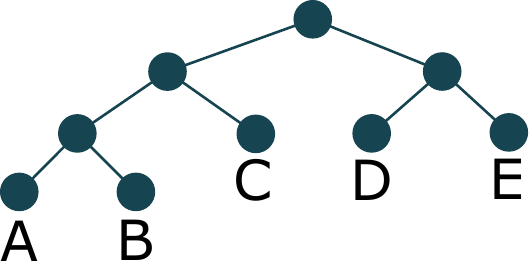}
    \caption{An example contraction tree with labeled leaves.}
    \label{fig:contree}
\end{figure}

Say we are given matrices \(M_1, M_2, \ldots, M_{n-1}, M_n\) with indices \(\{i_1, i_2\}, \{i_2, i_3\}, \ldots,  \{i_{n-1}, i_{n}\}, \{i_n, i_{n+1}\}\). The Matrix Chain Ordering Problem aims to find the optimal parenthesization of the matrix chain expression 
\begin{align*}
    M_1 \ast M_2 \ast \cdots \ast M_{n-1} \ast M_n
\end{align*}
which minimizes the cost of computing this product \cite{Cormen09}. Note that this setup assumes each matrix shares indices only with its neighbors. A given parenthesization can be expressed as a binary tree with leaves labeled with matrices. For example, the binary tree in Fig. \ref{fig:contree} corresponds to the parenthesized matrix chain
\begin{align*}
    ((A \ast B) \ast C) \ast (D \ast E).
\end{align*}

We define tensors analogously. A tensor \(X\) is an N-dimensional array of scalar numbers from some field \(\mathbb{F}\) with indices \(I = \{i_1, \ldots, i_N\}\) (for other perspectives, see \cite{biamonte_nutshell}). 

For a tensor \(X\) and an index \(i\), we say that \(X\) and \(i\) are incident to one another if \(i\) is an index of \(X\). We define \(\incident(X)\) as the set of indices incident to \(X\) and \(\incident(i)\) as the set of tensors incident to \(i\). The size of a tensor \(X\) is given by 
\begin{align*}
\tsize(X) &= \prod_{i \in \incident(X)} \tsize(i)
\end{align*}

Two tensors \(X_1\) with indices \(I_1\) and \(X_2\) with indices \(I_2\) can  similarly be combined using an operation known as tensor contraction. The output \(X_1 \ast X_2\) will have indices \(I_1 \triangle I_2\), where \(\triangle\) denotes symmetric difference. That is, the output has indices equal to the union of the input indices, with the intersection removed. \(I_1\triangle I_2\) are known as output indices, while \(I_1 \cap I_2\) are known as the shared indices. The number of scalar operations required to perform a tensor contraction is given by 
\begin{align*}
    \ops(X, Y) &= \tsize(X \ast Y)\shared(X, Y)
\end{align*}
where 
\begin{align*}
   \shared(X,Y) &= \prod_{i \in S} \tsize(i), & S &=  \incident(X)\cap \incident(Y).
\end{align*}

A set of tensors \(\mathcal{T} = \{X_1, \ldots, X_n\}\) with indices \[\mathcal{I} = \bigcup_{X \in \mathcal{T}} \incident(X)\] is known as a tensor network. Defining \(\mathcal{T}^{(0)} = \mathcal{T}\), a tensor network contraction iteratively picks \(X^{(k)},Y^{(k)} \in \mathcal{T}^{(k)}\) to produce \[\mathcal{T}^{(k+1)} = \left(\mathcal{T}^{(k)} \setminus \left\{X^{(k)}, Y^{(k)}\right\}\right) \cup  \left\{X^{(k)} \ast Y^{(k)}\right\},\]
terminating when \(\lvert \mathcal{T}^{(k)}\rvert = 1\). A contraction order used for this tensor network contraction can again be represented as a binary tree \cite{ogorman_parameterized}. For example, the tree shown in Fig. \ref{fig:contree} corresponds to the network contraction
\begin{align*}
    \mathcal{T}^{(0)} &= \{A, B, C, D, E\}\\
    \mathcal{T}^{(1)} &= \{A \ast B, C, D, E\}\\
    \mathcal{T}^{(2)} &= \{(A \ast B) \ast C, D, E\}\\
    \mathcal{T}^{(3)} &= \{(A \ast B) \ast C, D \ast E\}\\
    \mathcal{T}^{(4)} &= \{((A \ast B) \ast C) \ast (D \ast E)\}.
\end{align*}

Contraction of a tensor network is \(\#\)P-Hard in the general case \cite{ogorman_parameterized}. However, many tensor networks arising in practice may be contracted efficiently given a good quality contraction order with respect to some cost function. The costs we consider in this paper are the sum of the number of scalar operations, maximum number of scalar operations, and maximum tensor size, which are defined as follows:
\begin{gather}
    \sum_k \ops\left(X^{(k)}, Y^{(k)}\right),\label{eq:totalcost}\\
    \max_k \ops\left(X^{(k)}, Y^{(k)}\right), \qquad \max_k \tsize\left(X^{(k)} \ast Y^{(k)}\right).\label{eq:maxcost}
\end{gather} 
The problem of finding an optimal contraction order is itself NP-Hard \cite{ogorman_parameterized}. 

While minimizing the number of scalar operations required would seem to be more useful than minimizing the maximum number of operations for any contraction, the latter is actually useful in a parallel setting where many contractions can be performed simultaneously \cite{lee_alloc}.

\subsection{Quantum Circuits and QAOA}


An overview to quantum circuits and quantum computing can be found here \cite{watrous_circuit}.
A particular type of quantum circuit which has garnered great deal of research interest in recent years is the Quantum Approximate Optimization Algorithm ansatz, particularly those developed to target the graph theoretic MaxCut problem \cite{farhi2014quantum}. QAOA is a variational quantum-classical algorithm inspired by the adiabatic evolution principle. It is essentially a quantum annealer with a finite number of steps \(p\). The quality of the final solution increases with \(p\), as does the complexity of the circuit.

In quantum circuit simulation, it is possible to represent a quantum circuit as a high dimensional tensor over the complex numbers \cite{yanofsky_intro} or, equivalently, as a tensor network\cite{markov2008simulating}. In this perspective, each quantum gate is associated with a tensor \(X\) with indices corresponding to the inputs and outputs of the gate. Two tensors in such a network will share an index only if it corresponds to a value that has been passed from one corresponding gate to the other. For the purposes of this paper, we will assume the size of every index in the network is 2.

With this construction, the costs described in Eq \ref{eq:maxcost} are related to older literature which achieved similar estimations on the computational cost of tensor network contractions via quantities called vertex congestion, edge congestion, and treewidth \cite{markov2008simulating, ogorman_parameterized}. 

In particular, vertex congestion and treewidth are equal to the log of the maximum number of operations needed for any contraction
\begin{align}\label{eq:vertex}
    \max_k \log_2\ops\left(X^{(k)}, Y^{(k)}\right),
\end{align}
while edge congestion is equal to the log of the maximum size of any intermediate tensor 
\begin{align}\label{eq:edge}
    \max_k \log_2\tsize\left(X^{(k)} \ast Y^{(k)}\right).
\end{align}

\subsection{Tensor Networks as Graphs}

 An undirected graph \(G = (V, E)\) with no loops and multi-edges, is a set \(V\) of vertices  and a set \(E \subseteq \binom{V}{2}\) of edges.
The weights on the edges of the graph are denoted by the weighting function \(w\colon E \to \Real_{\geq 0}\). For an edge $uv\in E$, $w(uv)$ denotes its weight. The set of incident edges of a vertex \(v\) is the set of edges containing \(v\), denoted \(\incident(v) = \{ e \in E \mid v \in e \}\). The degree of \(v\) is the number of edges incident to \(v\), \(\degree(v) = \lvert \incident(v)\rvert\). 

Say we have a tensor network \(\mathcal{T}\) with indices \(\mathcal{I}\) where \(\lvert \incident(i)\rvert = 2\) for all \(i \in \mathcal{I}\). Then \(\mathcal{T}\) admits a representation as a weighted, undirected graph \(G = (V, E)\). Let \(F\) be the function which assigns tensors to vertices and indices to edges, such that \(F(i) = \{ F(X), F(Y) \} \) if and only if \(i \in \incident(X) \cap \incident(Y)\). The weight function \(w\) is defined as \(w(F(i)) = \log_2\tsize(i)\).





A linear vertex ordering for a graph \(G = (V, E)\) is a bijective mapping \(\sigma\colon V \to \{1, \ldots, \lvert V\rvert\} \). A linear ordering problem on a graph generally aims to find a linear ordering which minimizes a given cost function, such as the \(p\)-Sum objective, which is defined as 
\begin{align*}
    \left(\sum_{uv\in E} w(uv)\lvert \sigma(u) - \sigma(v)\rvert^p\right)^{1/p}
\end{align*}
\cite{safro_ordering}. For the special case where \(p = 1\), this problem is known as the Minimum Linear Arrangement Problem \cite{safro_mla}.

\subsection{Multilevel algorithms and refinement} 

The multilevel approach is a big class of algorithms actively used in many different areas of scientific computing, optimization, and machine learning \cite{brandt:optstrat}. In the context of this work, we briefly describe a version of multilevel algorithms for graph optimization problems \cite{ron2011relaxation,cheval-mlpartcompar}. When the graph is large and a fast solution of an optimization problem is required for a specific application, it is often useful to compress the problem by (possibly nonlinearly) aggregating variables into, so called, coarse variables. This is done in a such way that a solution for each coarse variable can be effectively interpolated back to those variables that participated in the aggregation. The entire process of problem coarsening is performed gradually forming a multilevel hierarchy of coarse problems. Each next coarser problem approximates the previous finer problem from which it has been created. Thus, the number of variables at each level of this hierarchy is decreasing which allows to solve them faster that the original problem. When sufficiently small coarsest level is created, the best possible (often exact) solution is computed. The last stage of framework (called uncoarsening) is to gradually solve the problems at each level of coarseness by (1) interpolating the initial solution for the current fine level from the coarser level, and (2) refining the interpolated solution. 

If both coarsening and uncoarsening are computed locally (i.e., their complexity is linear in the number of variables) then the entire multilevel framework becomes of linear or nearly-linear complexity assuming that the number of variables at each level is decreasing within a factor of 1.5-2.5. It is important to mention that this approach is different than such one-shot compression approaches as truncated SVD. Such problems on graphs as partitioning \cite{abou2006multilevel,amg-sss12}, various linear orderings \cite{Hu:wavefront,SafroT11}, and community detection \cite{ushijima2019ml,inuwa2021multilevel} have benefited from the multilevel approaches to mention just few of them \cite{Walshaw2004}.

\section{Related Work}\label{sec:related}

\subsection{FlowCutter}

FlowCutter is used collectively to refer to an algorithm for computing small, balanced s-t cuts using Pareto optimization \cite{hamann_flowcutter}, as well as an algorithm utilizing the former to construct a tree decomposition of the input graph via recursive bisection \cite{strasser_flowcutter}, that is also a form of a multilevel algorithm.

FlowCutter was entered into heuristic track of the PACE 2017 Parameterized Algorithms and Computational Experiments Challenge  on minimizing treewidth, where it placed second as the only solver to find a solution for all test instances in the allotted 30 minutes \cite{strasser_flowcutter, pace2017}. Moreover, FlowCutter is known to be useful in finding high quality tensor network contraction \cite{dudek_contraction}. 

\subsection{Tamaki-2017}

The Tamaki-2017 solver is a tree decomposition solver with components submitted to the exact and heuristic tracks for PACE 2017 \cite{pace2017}. On  the heuristic track, this solver ranked first place, tending to achieve better results on smaller instances than the competitive approaches, while sometimes failing to find an answer in the allotted time for larger instances \cite{strasser_flowcutter}.

The Tamaki-2017 algorithm is based on positive-instance driven dynamic programming, and, similar to FlowCutter, operates recursively on minimal separators of the instance graph\cite{tamaki_positive}.


\subsection{Cotengra}

Cotengra is a library for the efficient contraction of tensor networks \cite{cotengra}. In their seminal paper on the subject, the authors introduce a method for directly constructing contraction trees based on recursive bisection \cite{gray_hyperoptimized}. By utilizing existing hypergraph partitioning solvers such as KaHyPar \cite{schlag2021high}, Cotengra is able to construct contraction trees utilizing the hypergraph structure of some inputs that arise in Quantum Circuit Simulation \cite{gray_hyperoptimized}. 
Cotengra also utilizes the Bayesian optimization strategy, varying the hyperparameters of the partitioning solver at various levels in order to account for a changing graph structure as the algorithm progresses \cite{gray_hyperoptimized}.


\subsection{cuTENSOR}

cuTENSOR is a tensor network CUDA library in development by NVidia, which aims to bring high performance tensor primatives to the GPU \cite{cutensor}. Because not much information about their specific method for determining contraction orders is publicly available at the time of writing, we will not be comparing to cuTENSOR at this time.

\section{Constructing a Contraction Tree}\label{sec:method}

Our approach is founded on the observation that a parenthesization of a matrix chain expression or a contraction order for a tensor network both admit representations as binary trees with labelled leaves.

It is useful to introduce a more direct tensor variant of the Matrix Chain Ordering Problem, known as the Tensor Chain Ordering Problem. Given tensors \(X_1, \ldots, X_n\), find an optimal parenthesization of the tensor chain expression
\begin{align*}
    X_1 \ast \cdots \ast X_n
\end{align*}
which minimizes one of the costs introduced in Eqs \ref{eq:totalcost},\ref{eq:maxcost}. Notably, indices are not restricted to immediate neighbors in the chain.

Given a binary tree \(T\) representing a contraction order \(O\) for a tensor network \(\mathcal{T}\), consider the linear ordering \(\sigma\) of tensors corresponding to a left to right traversal of the labelled leaves of \(T\). Then \(T\) could be considered a possible solution to the Tensor Chain Ordering Problem for the chain
\begin{align*}
    \sigma^{-1}(1) \ast \cdots \sigma^{-1}(n).
\end{align*}
Then finding a contraction order for a tensor network can actually be reframed as finding a linear ordering of the tensors in the network, then solving the Tensor Chain Order Problem.
\begin{align*}
    \min_{O} \cost(\mathcal{T}, O) &= \min_{\sigma}\min_{T} \cost(\mathcal{T}, \sigma, T). 
\end{align*}
We call solving the Tensor Chain Order Problem for a fixed linear order \(\sigma\) tree structure optimization. Notably, an optimal contraction order of the Tensor Chain Order Problem can be found deterministically in polynomial time as demonstrated in the following section. This decouples the actual computationally Hard task of finding an optimal order of the chain, from the more tractable task of actually constructing the tree.

\subsection{Tree Structure Optimization}

In this section, we introduce a simple dynamic programming solution to the Tensor Chain Order Problem.
Say we are given the tensor chain expression \(X_1 \ast \cdots \ast X_n\). Let \(X_{i,j} = X_i \ast X_{i+1} \ast \cdots \ast X_j\) for \(i < j\).

From this, we find the following recursion for computing the minimal edge congestion of any parenthesization of the tensor chain
\begin{align}
\begin{split}
    c(i, i) &= \tsize(X_i)\\
    c(i, j) &= \min_{i \leq k < j}\max \begin{cases}
    \log_2 \tsize(X_{i, k} \ast X_{k+1, j})\\
    c(i, k)\\
    c(k+1, j)
    \end{cases} 
\end{split}
\end{align}
as well as the minimal vertex congestion
\begin{align}
\begin{split}
    c(i, i) &= \tsize(X_i)\\
    c(i, j) &= \min_{i \leq k < j}\max \begin{cases}
    \log \ops(X_{i, k}, X_{k+1, j})\\
    c(i, k)\\
    c(k+1, j)
    \end{cases} 
\end{split}
\end{align}

It is beneficial to avoid attempting to calculate \(\shared(X_{i,k}, X_{k+1,j})\) at each iteration. Instead, we introduce the subroutine Alg. \ref{alg:between} which more efficiently calculates every value of shared as \(k\) is varied across the sequence. If \(d = \max_{X \in \mathcal{T}} \lvert \incident(X)\rvert\) is the maximum degree of any tensor in \(\mathcal{T}\), this can be done in \(O(d\lvert \mathcal{T}\rvert)\) time. 

\begin{algorithm}
\caption{CalcShared}\label{alg:between} 
\begin{algorithmic}
\REQUIRE  A tensor chain \(C = (X_1, \ldots, X_n)\), \(1 \leq i \leq j \leq n\)
\STATE \(shared \gets [0 \mid  \text{for } i \leq k < j ]\)
\STATE \((upper, lower) \gets (\{X_i, \ldots, X_{k-1}\}, \{X_{k+1},\ldots, X_j\})\)
\FOR{\(i \leq k < j\)}
\STATE \(new \gets sum(\tsize, \{ a \in \incident(X_i) \mid \incident(a) \cap upper \neq \emptyset\})\)
\STATE \(old \gets sum(\tsize, \{ b \in \incident(X_i) \mid \incident(b) \cap lower \neq \emptyset\})\) 
\STATE\(shared[k] \gets shared[k] +  new - old\)
\STATE\(shared[k+1] \gets shared[k]\)
\ENDFOR
\RETURN \(shared\)
\end{algorithmic}
\end{algorithm}

From these recursions, we introduce Alg. \ref{alg:alg}, a dynamic programming algorithm for computing the minimal possible size of the largest tensor for any contraction tree given an order. We make use of pruning in order to reduce the number of needed computations, as there is no need to evaluate the second subsequence in a split if the first has already exceeded the current best result. Performance can be further improved by making use of memoization to cache previous calls. Additional gains may be achieved by parallelizing the evaluation of disjoint subproblem calls, which is an ongoing project. Utilizing these techniques, the complexity of Alg. \ref{alg:alg} is \(O(\lvert E\rvert \lvert V\rvert ^ 2 + d\lvert V\rvert ^ 3)\).





\begin{algorithm}
\caption{Tree Structure Optimization}\label{alg:alg}
\begin{algorithmic}
\REQUIRE A tensor chain \(C = (X_1, \ldots, X_n)\), \(1 \leq i \leq j \leq n\)
\REQUIRE \(x \oplus y = x + y \text{ or } \max(x, y)\)
\IF{\(i == j\)}
    \RETURN \((\tsize(X_i), i)\)
\ENDIF
\STATE \(shared \gets CalcShared(C, i, j)\)
\STATE \(outsize \gets \tsize(X_{i, j})\)
\STATE \((c^*, t^*) \gets \infty, i\)
\FOR{\(i \leq k < j\)}
    \STATE \(c \gets COST(outsize, shared[k])\)
    \IF{\(c > c^*\)}
    \item
        continue
    \ENDIF
    \STATE \((c_l, l) \gets TreeStructureOptimization(C, i, k)\)
    \STATE \(c \gets c \oplus c_l\)
    \IF{\(c > c^*\)}
    \item continue
    \ENDIF
    \STATE \((c_r, r) \gets TreeStructureOptimization(C, k+1, j)\)
    \STATE \(c \gets c \oplus c_r\)
    \IF{\(c < c^*\)}
    \STATE \((c^*, t^*) \gets (c, (l, r))\)
    \ENDIF
\ENDFOR
\RETURN \((c^*, t^*)\)
\end{algorithmic}
\end{algorithm}

A variant for the classical total number of operations can be achieved simply by looking at the incremental sum of the left and right trees rather than the max.
\begin{align} 
\begin{split}
    c(i, i) &= \tsize(X_i)\\
    c(i, j) &= \min_{i \leq k < j}\ops(X_{i, k}, X_{k+1, j}) + c(i, k) + c(k+1, j)
\end{split}
\end{align}

\subsection{Heuristic Order Choices}


While we have introduced algorithms for constructing an optimal contraction tree given a particular order, we must now tackle the problem of actually choosing an order. We turn to a heuristic approach for solving this problem. In particular, we select an order which minimizes the sum total of lengths of stretched edges in the graph. This is known as the Minimal Linear Arrangement of the vertices in the graph, and is defined as 
\begin{align*}
    \sum_{uv \in E} w(uv)\lvert \sigma(u) - \sigma(v)\rvert
\end{align*}

This is motivated by the property of matrix chains sharing indices only with their immediate neighbors. An order which minimizes MLA discourages long stretched edges on the chain. We have empirically observed this to be effective in reducing the congestion of the resulting contraction tree, which will be discussed further in the Results section.


In order to calculate an order which effectively minimizes MLA, we utilize the LinearOrdering.jl package, a Julia package for multilevel linear ordering that is currently in development by the authors. This package utilizes a volume based coarsening strategy as well as node by node minimization in order to find a high quality arrangement \cite{safro_mla}. The package can be found at \url{https://github.com/cameton/LinearOrdering.jl}.

\section{Experimental Results}\label{sec:results}


We evaluate tree structure optimization against existing state of the art algorithms Tamaki-2017, FlowCutter, and Cotengra on a collection of randomly generated QAOA circuits. To generate test results for tree structure optimization, we find an order for the graph with a good Minimum Linear Arrangement objective, then run tree structure optimization with the relevant cost. This is done repeatedly while varying the random seed for a set amount of time, returning the best result at the end. In most cases we observe a significant improvement in the quality of results given a comparable running time (that is usually negligible in comparison to the quantum simulation time itself).

For our tests, for a fixed depth \(p\) and degree \(d\), we construct a \(p\) layer QAOA ansatz circuit targeting the Max Cut problem on a random 32 vertex \(d\)-regular graph using the QTensor library\cite{qtensor}. Each circuit will have a number of qubits equal to the number of vertices in the given graph. For each combination of \(d = 3, 4, 5\) and QAOA depth \(p = 2, 3, 4, 5\), we generate 10 random circuits to form our test set. The source code and generated circuits are available at \url{https://github.com/cameton/HPEC2022_ContractionTrees}.
In doing so, we sample a variety of circuits with a varying level of connectedness and depth, which we use as a proxy for the complexity of the circuit; such circuits are commonly used in evaluations of contraction order solvers for quantum circuits as in \cite{gray_hyperoptimized}.


\begin{figure*}[ht]
    \centering
     \begin{subfigure}{0.475\textwidth}
         \centering
         \includegraphics[width=\textwidth]{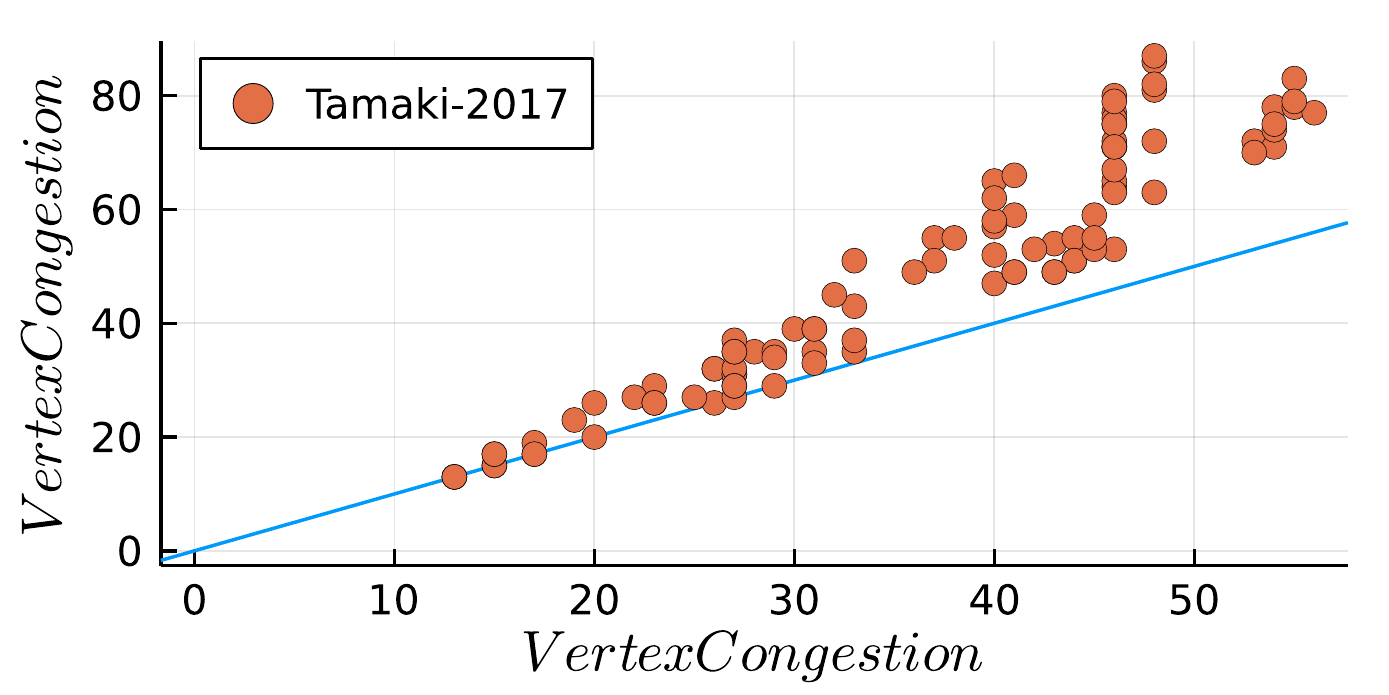}
     \end{subfigure}
     \hfill
    \begin{subfigure}{0.475\textwidth}
         \centering
         \includegraphics[width=\textwidth]{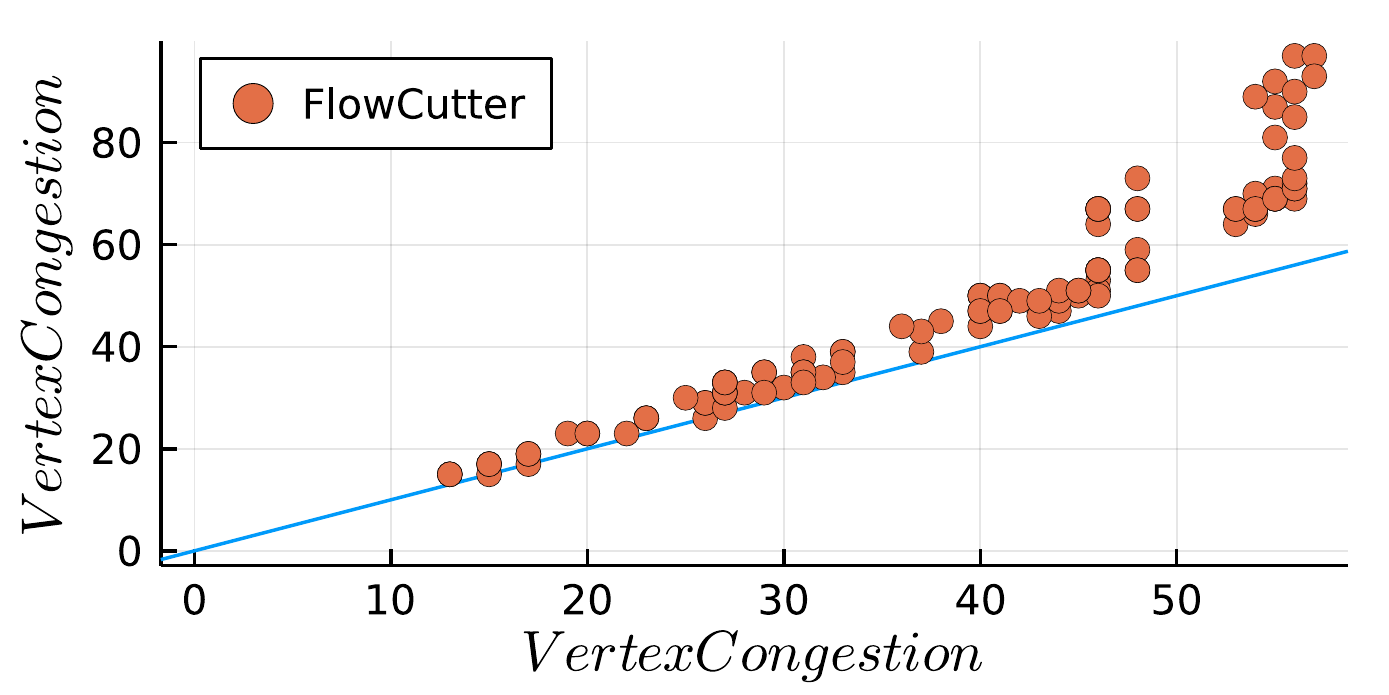}
     \end{subfigure}
    \caption{Vertex Congestion achieved by Tamaki-2017  (left) and FlowCutter (right) plotted against the vertex congestion found by our tree structure optimization on a Minimum Linear Arrangement order. The blue line represents the threshold at which both solvers achieved the same objective value. Points above this threshold indicate that tree structure optimization found a superior value.}
    \label{fig:congestion}
\end{figure*}

\subsection{Vertex Congestion}

We compare the vertex congestion (Eq. \ref{eq:vertex}) of the solution found by our tree structure optimizer for the test set against that found by Tamaki-2017 and FlowCutter by finding the tree decomposition of the line graph. As Cotengra has functionality to minimize edge congestion and number of FLOPS, it will be evaluated separately. For these experiments, we ran each of these optimizers for 5 seconds. Results for this test are given in Fig. \ref{fig:congestion}.



Our tree structure optimization based approach performed strictly better than Flowcutter on 96.4\% of points and better than Tamaki on 90.0\% of points, while never achieving a worse cost. What's more, the margin of improvement increases significantly as the vertex congestion of the circuit increases.

As vertex and edge congestion represent the log of the maximum size and maximum number of operations, even small improvements in congestion lead to significant gains. As such, having such a large improvement on so many points represents advantage of our solver over competitive solvers run for the same amount of time.

\subsection{Edge Congestion and Operation Count} 

\begin{figure*}[ht]
    \centering
     \begin{subfigure}{0.45\textwidth}
         \centering
         \includegraphics[width=\textwidth]{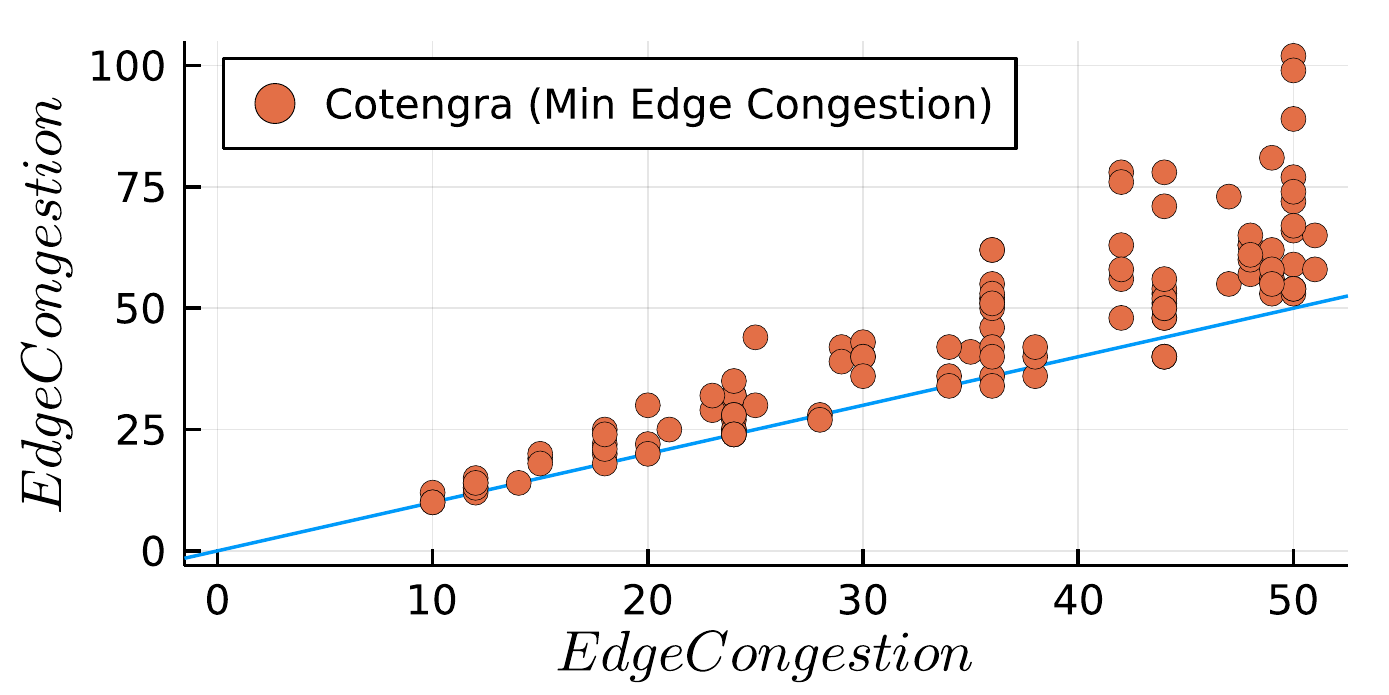}
     \end{subfigure}
     \hfill
     \begin{subfigure}{0.45\textwidth}
         \centering
         \includegraphics[width=\textwidth]{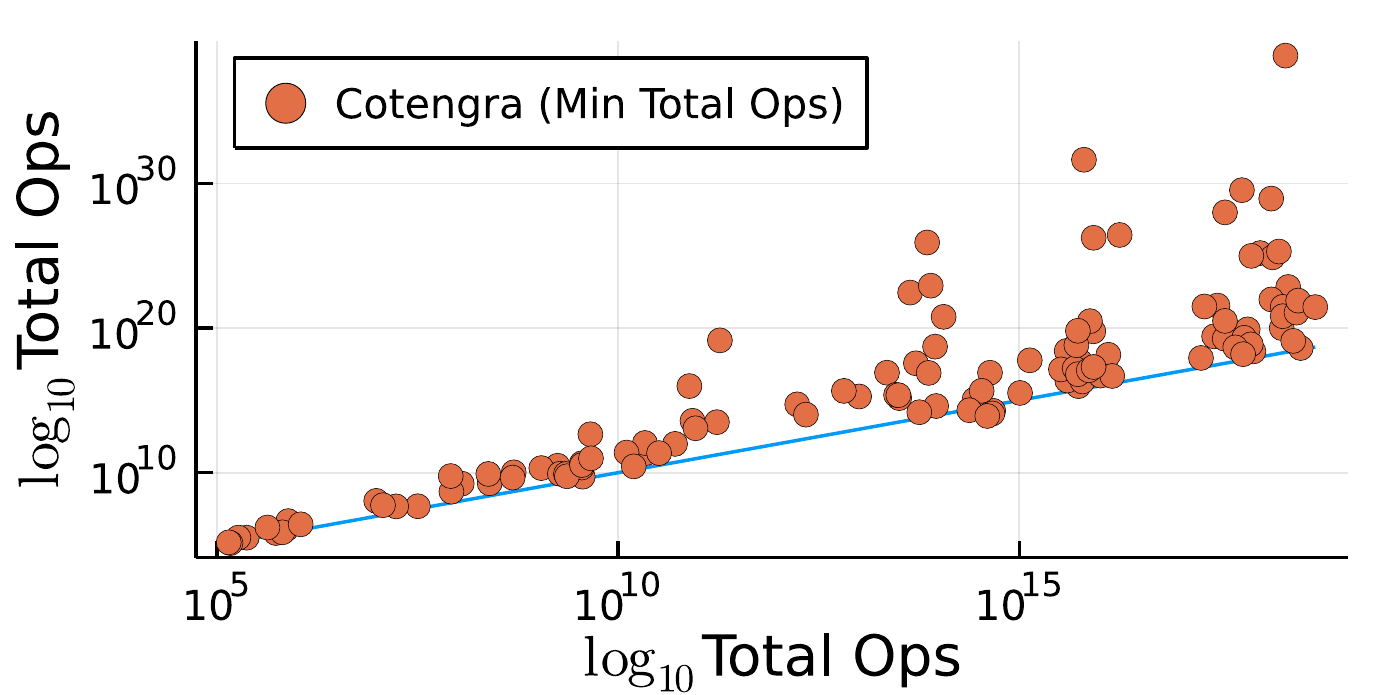}
     \end{subfigure}
    \caption{Edge congestion of Cotengra minimizing edge congestion against tree structure optimization (left) and Cotengra minimizing operation count against tree structure optimization on a log-log scale (right). The blue line represents the threshold at which both solvers achieved the same objective value. Points above this threshold indicate that tree structure optimization found a superior value.}
    \label{fig:cotengra}
\end{figure*}

We compare the edge congestion (Eq. \ref{eq:edge}) and estimated number of operations (Eq. \ref{eq:totalcost}) of the solution found by our tree structure optimizer on the test set against those found by Cotengra. For these tests, Cotengra was run for 5 seconds with default settings and the KaHyPar partitioner backend. Two tests were run using Cotengra: one minimizing edge congestion and one minimizing the total number of scalar operations. Results are given in Fig. \ref{fig:cotengra}.

From Fig. \ref{fig:cotengra}, we see that tree structure optimization is highly competitive with Cotengra, in every case finding a solution which is close to or better than Cotengra in terms of both estimated number of FLOPS and edge congestion.


Our solver bests Cotengra on a full 95\% of points when minimizing total number of operations, and on 84.1\% of points when minimizing edge congestion. This once again demonstrates the superiority of our solver for comparatively short run times.

\subsection{Varying the Number of Qubits}

We may also wish to compare these solvers as we vary the number of vertices in the input graph, by extension varying the number of qubits. To avoid redundancy, we restrict our comparison for this section to Cotengra and Total FLOPS alone. For these tests, we construct a \(2\) layer QAOA ansatz circuit targeting the Max Cut problem for random \(3\)-regular graphs with \(32, 64, 96\) and \(128\) vertices. For each size, we generate 10 random circuits. Because each circuit has a number of qubits equal to the size of the input graph, we can examine circuits of a similar complexity across a variety of different qubit counts. For these tests, we ran both Cotengra and the ordering solver for 10 seconds for each input.

\begin{figure}[ht]
     \centering
     \includegraphics[width=0.45\textwidth]{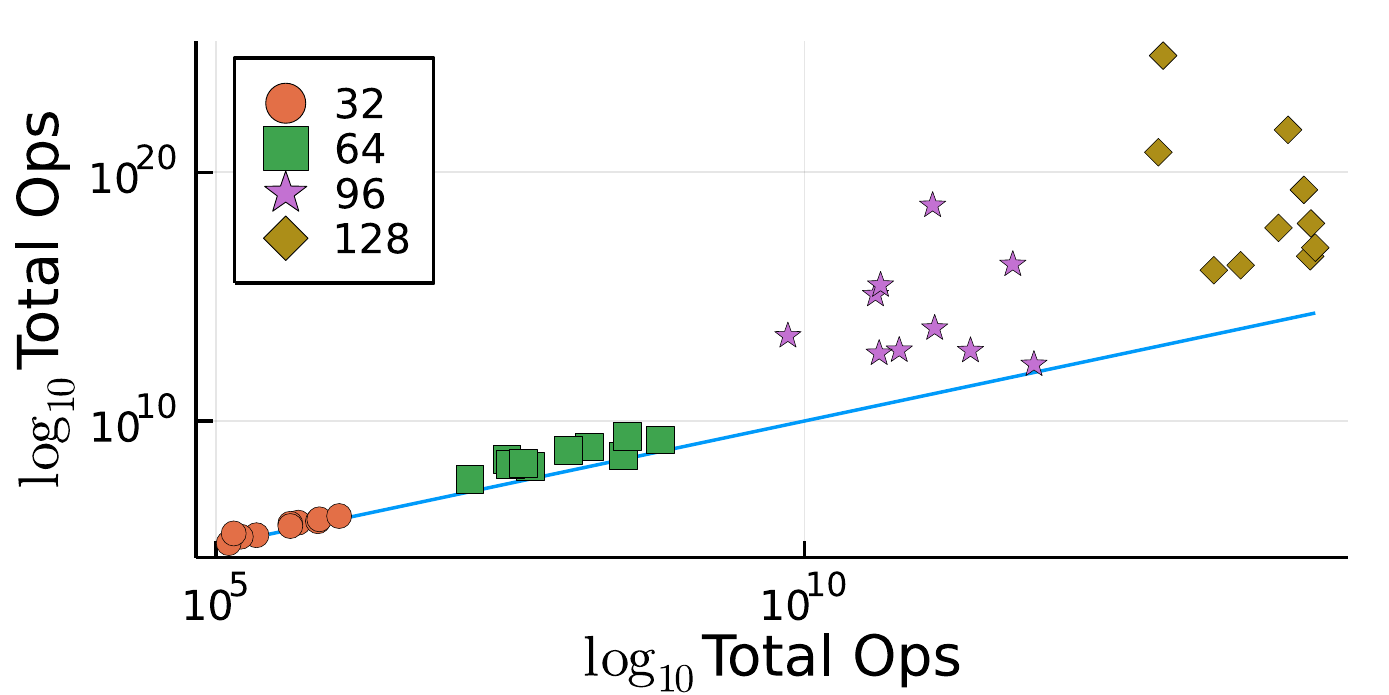}
    \caption{Cotengra minimizing operation count against tree structure optimization on a log-log scale for circuits with 32, 64, 96, and 128 qubits. The blue line represents the threshold at which both solvers achieved the same objective value. Points above this threshold indicate that tree structure optimization found a superior value.}
    \label{fig:cotengra_qubits}
\end{figure}

The results of this test are shown in Fig. \ref{fig:cotengra_qubits}. Averaging over the 32 qubit circuits, our tree structure optimization based approach produced results around 28.6\% better than Cotengra, improving to 74.1\% when averaging over the 64 qubit circuits. For the 96 and 128 qubit circuits, nearly every result produced by our tree structure optimization approach is several orders of magnitude better than Cotengra for the 10 second running time.

\section{Future Work}\label{sec:future}

\subsection{Width Refinement}

The current refinement strategy used in this paper minimizes Minimum Linear Arrangement by making local changes to the order at each level of the algorithm \cite{safro_ordering}. While we have shown thus far that Minimum Linear Arrangement works as a effective heuristic for the tested circuits, we may be able to achieve further improvements through the development of refinement strategies which directly target the width of the order. That is, we would like the ability to make local changes which serve to reduce the width of the overall order, potentially achieving better results than the more general Minimum Linear Arrangement heuristic. 

\subsection{Performance \& Scalability}

The algorithms introduced here for tree structure optimization are currently not parallelized. The development of a tree structure optimization algorithm which takes advantage of the graph structure for improved performance  is vital to the application of this technique to fields outside of quantum circuit simulation, such as in large scale shortest path acceleration\cite{hamann_flowcutter}.

Although the multilevel algorithm component is already fast, it will be valuable to develop a strongly parallelized version of these algorithms which may be better suited for a high performance computing environment, particularly where such hardware is being used for circuit simulation already.

\subsection{Higher Order Structures}

Thus far we have primarily considered tensor networks representable as normal undirected graphs. In this context, we associate each index in the circuit with a corresponding edge in the graph. For some quantum circuits, a possible cost-saving technique is to reconsider certain collections of indices as hyperindices, in essence a single index connecting multiple gates \cite{lykov2021importance}. This introduces a hypergraph structure to our tensor network representation, a higher order generalization of undirected graphs.

As it stands, our work has yet to be extended to accommodate these higher order structures, something which Cotengra is currently able to handle by making use of existing hypergraph partitioning software \cite{gray_hyperoptimized}. In particular, attention needs to be paid to how the hypergraph structure may inform the coarsening portion of the ordering problem. For example, this can be done by evaluating various kernels and similarity measures on high order structures such as in \cite{fronczak2002higher,shaydulin2018aggregative,bai2012jensen,shaydulin2017relaxation}. Moreover,  this necessitates a tree structure optimization algorithm for higher order structures.

\section{Conclusion}\label{sec:conclusion}

Accelerating quantum circuit simulation is a critical problem not only for demonstrating quantum advantage but also for hybrid quantum-classical algorithms which require computationally heavy parameter optimization backend on the classical machine. In this paper, we presented novel algorithms for constructing an optimal contraction tree from a given order. We introduced a multilevel solver for the Minimum Linear Arrangement problem on graphs
and demonstrated its applicability as a heuristic for providing orderings for contraction trees. We compared the performance of our solver against state-of-the-art industry solvers Tamaki-2017, FlowCutter, and Cotengra on a collection of randomly generated QAOA circuits. We have shown that our method achieves results that are superior by orders of magnitude on some instances to competitors when run for a comparable amount of time. There is work under way to produce higher quality orders for producing contraction trees, and on improving the speed of tree structure optimization.

\section*{Acknowledgements}

This material is based upon work supported by the
Defense Advanced Research Projects Agency (DARPA) ONISQ program 
under Contract No. HR001120C0068. This research was partially supported by NSF, under award number 2122793. Y.A.’s and D.L.'s work at Argonne National Laboratory was partially supported by the U.S. Department of Energy, Office of Science, under contract DE-AC02-06CH11357.

\bibliographystyle{plain}
\bibliography{onesum,fullbib}

\end{document}